\documentclass[a4paper,12pt]{article}

\usepackage{graphicx}
\usepackage{amsmath}
\usepackage{amssymb}
\usepackage{hyperref}
\usepackage{tabularx}
\newcolumntype{M}{>{$}c<{$}}
\usepackage[matrix,arrow,curve]{xy}

\numberwithin{equation}{section} \numberwithin{figure}{section}
\numberwithin{table}{section}

\def\papertitlepage{\baselineskip 3.5ex\thispagestyle{empty}}
\def\Title#1{\baselineskip 1cm \vspace{1.5cm}%
  \begin{center}{\Large\bf #1}\end{center}\vspace{0.5cm}}
\def\Authors#1{\begin{center}\renewcommand{\thefootnote}{\fnsymbol{footnote}}{\it #1}\end{center}}
\def\Abstract{\vspace{1.0cm}%
  \begin{center}{\large\bf Abstract}\end{center}}

\renewenvironment{thebibliography}{\pagebreak[3]\par\vspace{0.6em}
\begin{flushleft}{\large \bf References}\end{flushleft}
\vspace{-1.0em}

\begin{enumerate}\if@twocolumn\baselineskip=0.6em\itemsep -0.2em
\else\itemsep -0.2em\fi\labelsep 0.1em}{\end{enumerate} }

\DeclareMathDelimiter{\lcolon}{\mathopen}{operators}{"3A}{largesymbols}{"3A}
\DeclareMathDelimiter{\rcolon}{\mathclose}{operators}{"3A}{largesymbols}{"3A}

\def\+{\!\!+\!\!}

\def\dynkin(#1){(#1)}

\def\bra<#1|{\langle#1|}
\def\ket|#1>{|#1\rangle}
\def\braket<#1|#2>{\langle#1|#2\rangle}
\def\llangle{\langle\!\langle}
\def\rrangle{\rangle\!\rangle}
\def\bbra<#1|{\llangle#1|}
\def\kket|#1>{|#1\rrangle}
\def\bbraket<#1|#2>{\llangle#1|#2\rrangle}

\begin{document}
{\papertitlepage \vspace*{0cm} {\hfill
\begin{minipage}{4.2cm}
IFT-P. 2009\par\noindent September, 2009
\end{minipage}}
\Title{The Tachyon Potential in the Sliver Frame}
\Authors{{\sc E.~Aldo~Arroyo${}$\footnote{\tt
aldohep@ift.unesp.br}}
\\
${}$
Instituto de F\'{i}sica Te\'{o}rica, UNESP - Universidade Estadual Paulista \\[-2ex]
Caixa Postal 70532-2 \\ 01156-970 S\~{a}o Paulo, SP, Brasil
\\
${}$ }

} 

\vskip-\baselineskip
{\baselineskip .5cm \Abstract We evaluate the tachyon potential in
the Schnabl gauge through off-shell computations in the sliver
frame. As an application of the results of our computations, we
provide a strong evidence that Schnabl's analytic solution for
tachyon condensation in open string field theory represents a
saddle point configuration of the full tachyon potential.
Additionally we verify that Schnabl's analytic solution lies on
the minimum of the effective tachyon potential.

\newpage
\setcounter{footnote}{0}
\tableofcontents

\section{Introduction}

Schnabl's \cite{Schnabl:2005gv} description of an analytic
solution for tachyon condensation has sparked renewed interest in
string field theory in the last few years. The study of open
string tachyon condensation on unstable branes in bosonic and
superstring theory is interesting, since it involves three
important conjectures made by Ashoke Sen
\cite{Sen:1999mh,Sen:1999xm}. The first conjecture related the
height of the tachyon potential at the true minimum to the tension
of the D-brane; the second conjecture predicted existence of lump
solutions with correct tensions, which describe lower dimensional
D-branes; and the third conjecture stated that there are no
physical excitations around the true minimum \footnote{There are
similar conjectures for the open string tachyon on a non-BPS
D-brane and the tachyon living on the brane-antibrane pair
\cite{Sen:1998ii,Sen:1998sm,Sen:1999mg,Bai:2005jr,Bagchi:2008et}.}.
Witten's cubic or Chern-Simons open string field theory
\cite{Witten:1985cc} has provided precise quantitative tests of
these conjectures
\cite{Kostelecky:1988ta,Kostelecky:1989nt,Kostelecky:1995qk,Sen:1999nx,Moeller:2000xv,Taylor:2002fy,Gaiotto:2002wy,
Harvey:2000tv,de Mello
Koch:2000ie,Moeller:2000jy,Ellwood:2001py,Ellwood:2001ig,Giusto:2003wc,Ellwood:2006ba,Kwon:2007mh}.

Since string field theory corresponds to second-quantized string
theory, a point in its classical configuration space corresponds
to a specific quantum state of the first quantized string theory.
As shown in ref. \cite{Witten:1985cc}, in order to describe
a gauge invariant open string field theory we must include the
full Hilbert space of states of the first quantized open string
theory, including the $b$ and $c$ ghost fields. Witten's
formulation of open string field theory is based on the following
Chern-Simons action
\begin{eqnarray}
\label{accionx1} S=-\frac{1}{g^2}\Big[ \frac{1}{2} \langle
\Phi,Q_B \Phi \rangle +\frac{1}{3} \langle \Phi,\Phi*\Phi \rangle
\Big] \, ,
\end{eqnarray}
where $Q_B$ is the BRST operator of bosonic string theory, $*$
stands for Witten's star product, and the inner product $\langle
\cdot , \cdot \rangle $ is the standard BPZ inner product. The
string field $\Phi$ belongs to the full Hilbert space of the first
quantized open string theory. The action has gauge invariance
$\delta \Phi = Q_B \Lambda+\Phi*\Lambda-\Lambda*\Phi$.

The tachyon potential in Witten's cubic open string field theory
has been computed numerically by an approximation scheme called
level truncation
\cite{Kostelecky:1988ta,Kostelecky:1989nt,Kostelecky:1995qk,Sen:1999nx,Moeller:2000xv,Taylor:2002fy,Gaiotto:2002wy}.
This method is rather similar to the variational method in quantum
mechanics and it was first used by Kostelecky and Samuel. The
scheme is based on the realization that by truncating the string
field $\Phi$ to its low lying modes (keeping only the Fock states
with $L_0<h$), one obtains an approximation that gets more
accurate as the level $h$ is increased. Therefore,
the string field was traditionally expanded in the so-called Virasoro basis of
$L_0$ eigenstates. However, it is well known that in this
basis calculations involving the cubic interaction term becomes
cumbersome and the three-string vertex that defines the
star product in the string field algebra $\Phi_1*\Phi_2$ is
complicated
\cite{Gross:1986ia,Gross:1986fk,LeClair:1988sp,LeClair:1988sj}. We
can overcome these technical issues (related to the definition of
the star product) by using a new coordinate system
\cite{Schnabl:2005gv}.

The open string worldsheet is usually parameterized by a complex
strip coordinate $w=\sigma + i \tau$, $\sigma \in [0,\pi]$, or by
$z=-e^{-iw}$, which takes values on the upper half plane. As shown
in \cite{Schnabl:2005gv}, the gluing conditions entering into the
geometrical definition of the star product simplify if one uses
another coordinate system, $\tilde z = \arctan z$, in which the
upper half plane looks as a semi-infinite cylinder of
circumference $\pi$. In this new coordinate system, which we will
henceforth call the \textit{sliver frame}, it is possible to write
down simple, closed expressions for arbitrary star products within
the subalgebra generated by Fock space states. Elements of this
subalgebra are known in the literature
\cite{Rastelli:2000iu,David:2001xu,Schnabl:2002gg} as wedge states
with insertions.

The simplicity of the definition of the star product in the sliver
frame allows us to solve analytically the string field equation of
motion \cite{Schnabl:2005gv}
\begin{eqnarray}
\label{equamo1} Q_B \Phi + \Phi*\Phi =0 \, .
\end{eqnarray}
Schnabl's analytic solution $\Phi\equiv\Psi$ was obtained by
expanding the string field $\Psi$ in a basis of ${\cal L}_0$
eigenstates, where ${\cal L}_0$ is the zero mode of the worldsheet
energy momentum tensor $T_{\tilde z \tilde z}$ in the $\tilde z$
coordinate. By a conformal transformation it can be written as
\begin{eqnarray}
\mathcal{L}_0= \oint \frac{d z}{2 \pi i} (1+z^2) \arctan z \, T_{
z z}(z)=L_0+\sum_{k=1}^{\infty} \frac{2(-1)^{k+1}}{4k^2-1}L_{2k}
\; ,
\end{eqnarray}
where the $L_n$'s are the ordinary Virasoro generators with zero
central charge $c=0$ of the total matter and ghost conformal field
theory. The coefficients of the ${\cal L}_0$ level expansion of
the string field $\Psi$ are obtained by plugging $\Psi$ into the
equation of motion. Remarkably, imposing the Schnabl gauge
condition ${\cal B}_0 \Psi=0$ \footnote{$\mathcal{B}_0$ is the
zero mode of the $b$ ghost in the $\tilde z$ coordinate, which can
be defined by a conformal transformation in a similar manner as
${\cal L}_0$.} and truncating the equation of motion (but not the
string field) to the subset of states up to some maximal ${\cal
L}_0$ eigenvalue lead to a system of algebraic equations for the
coefficients, which can be solved exactly level by level. It was
shown that the analytic solution $\Psi$ reproduces the desired
value for the normalized vacuum energy predicted
from Sen's first conjecture
\cite{Okawa:2006vm,Fuchs:2006hw,Takahashi:2007du,Aref'eva:2009ac,Arroyo:2009ec,Erler:2009uj}
\begin{eqnarray}
\label{conjec1} 2 \pi^2 \Big[\frac{1}{2} \langle \Psi,Q_B \Psi
\rangle +\frac{1}{3} \langle \Psi,\Psi*\Psi \rangle  \Big]= -1 \,
.
\end{eqnarray}

To date, it is still an open question how to construct an explicit
gauge transformation to prove the equivalence between Schnabl's
analytic solution and the numerical solution found in the level
truncation scheme in the Siegel gauge. These two solutions are
believed to be the same tachyon vacuum solution. Evidence which
supports this statement is given by evaluating gauge invariants
quantities on these solutions, namely, the vacuum energy and the
gauge invariant overlap
\cite{Ellwood:2008jh,Kawano:2008ry,Kawano:2008jv,Kiermaier:2008qu,Kishimoto:2009cz}.
The computation of these gauge invariant quantities does not tell
us much about the type of configuration associated to the
solutions. Since, in general, a solution to the equation of motion
corresponds to extremal configurations, at first sight we do not
know that the solution will correspond to a minimum, maximum or
saddle point configuration of the theory. In the case of string
field theory a direct way to find the kind of configuration
associated with the solution is to compute the off-shell tachyon
potential. In this paper, we compute the tachyon potential in the
so-called sliver frame. By extremizing this potential, we search
for extremal configurations, and remarkably it turns out that
Schnabl's analytic solution for tachyon condensation represents a
saddle point configuration of the full tachyon potential.

Regarding to the effective tachyon potential, let us point out
that this potential is non-unique. In general we can compute the
effective tachyon potential as follows. By decomposing the string
field as $\Phi=t \mathcal{T} + \chi$, where $\mathcal{T}$ is the
tachyonic part of the string field (the zero momentum tachyon
state), while $\chi$ is an arbitrary string field which belongs to
the gauge fixed Hilbert space linearly independent of the first
term $\mathcal{T}$. To obtain the effective tachyon potential, we
must integrate out the string field $\chi$, this is done by
inserting the string field $\Phi$ into the action, solving the
equation of motion for $\chi$ and plugging back to the action. The
resulting expression, as a function of the single variable $t$ is
the effective potential. Therefore we see that the effective
potential computed in this way is non-unique since it depends on
the choice of an specific gauge to fix the string field $\Phi$ and
the choice of $\mathcal{T}$. For instance, traditionally
$\mathcal{T}$ is taken to be $c_1 |0\rangle$ and the gauge used is
the Siegel gauge $b_0 \Phi=0$.

Usually when we compute the effective tachyon potential in the
Siegel gauge \cite{Moeller:2000xv,Taylor:2002fy,Gaiotto:2002wy},
the fields which are integrated out correspond to the perturbative
Fock space of states with mass greater than the tachyon mass. In
the case of the Schnabl gauge, to find the effective tachyon
potential in the sliver frame, instead of integrating out fields
in the state space used in the Siegel gauge, we integrate out
fields with $\mathcal{L}_0$ eigenvalue greater than $-1$, which
corresponds to the $\mathcal{L}_0$ eigenvalue of the tachyon state
$\tilde c_1 |0\rangle$. This means that the effective tachyon
potential we compute in the sliver frame is different from the old
effective tachyon potential computed in the Siegel gauge.  As an
application of the results of our computations, we address the
question of whether Schnabl's analytic solution corresponds to a
minimum configuration of the effective tachyon potential, and we
find that this is indeed the case.

We have chosen $\mathcal{T}=\tilde c_1 |0\rangle$ as the tachyonic
part of the string field since we consider this choice to be the
most natural one from the perspective of Schnabl's coordinates
(the sliver frame). Choosing insertions on other wedge states does
not seem to be natural, except for insertions over the sliver or
the identity $\tilde c_1 |\infty\rangle$, $\tilde c_1
|\mathcal{I}\rangle$, nevertheless both of these options are
singular
\cite{Kluson:2002ex,Kishimoto:2001de,Takahashi:2002ez,Rastelli:2001jb}.
We believe that any choice other than $\mathcal{T}=\tilde c_1
|0\rangle$ would require some motivation. One of the main
motivations for computing the tachyon potential is to investigate
its branch structure, for example, to see if the perturbative and
non-perturbative vacua are connected on the same branch of the
potential, to see if there are other branches with other
solutions, and to understand the off-shell limitations of the
gauge-choice. If we choose $\mathcal{T}$ to be the tachyon vacuum
solution, the resulting potential is less interesting in this
respect. From a technical point of view our choice
$\mathcal{T}=\tilde c_1 |0\rangle$ appears to be the less involved
one.

Another remark we would like to comment is related to the choice
of basis for integrating out the remaining string field $\chi$.
The effective tachyon potential should not depend on the choice of
basis even if the basis are related by a somewhat singular
transformation like the transformation between the $L_0$ with the
$\mathcal{L}_0$ basis. This statement should be true provided that
we can manage the singularity by a suitable regularization
prescription. For instance, the use of Pad\'{e} resummation
techniques would eventually be needed
\cite{Takahashi:2007du,Arroyo:2009ec}.

This paper is organized as follows. In section 2, we introduce
Witten's formulation of open bosonic string field theory. After
writing down the form of the cubic action, we define the two- and
three-string interaction vertex. To evaluate these interaction
vertices, we use CFT correlators defined on the sliver frame. In
section 3, we study the structure of the tachyon potential in some
detail using the $\mathcal{L}_0$ level expansion of the string
field in the Schnabl gauge. By extremizing the potential, we
provide a strong evidence that Schnabl's analytic solution
corresponds to a saddle point configuration of the theory. A
summary and further directions of exploration are given in section
4.

\section{Open bosonic string field theory revisited}
In this section, we are going to review briefly some aspects of
Witten's cubic open string field theory which will be relevant to
the purposes of this paper.

\subsection{Witten's string field theory}
Witten's formulation of open string field theory is
axiomatic. The space of string fields $\mathcal{H}$ is taken to be an associative noncommutative algebra provided with a $\mathbf{Z}_2$
grading and a $*$-multiplication operation on $\mathcal{H}$.
The multiplication law $*$ satisfies the property that the
$\mathbf{Z}_2$ degree of the product $a * a'$ of two elements $a$,
$a'$ $\in$ $\mathcal{H}$ is $(-1)^a(-1)^{a'}$, where $(-1)^a$ is
the $\mathbf{Z}_2$ degree of $a$. There exists an odd derivation
$Q$ acting on $\mathcal{H}$ as $Q(a* a') = Q(a) * a' + (-1)^a a *
Q(a')$. $Q$ is also required to be nilpotent: $Q^2 = 0$. These
properties remind us of the BRST operator $Q_B$. The final
ingredient is the integration, which maps $a$ $\in$ $\mathcal{H}$
to a complex number $\int a$ $\in$ $\mathbb{C}$. This operation is
linear, $\int(a+a') = \int a+\int a'$, and satisfies $\int (a*a')
= (-1)^{aa'} \int (a' *a)$ where $(-1)^{aa'}$ is defined to be
$-1$ only if both $a$ and $a'$ are odd elements of $\mathcal{H}$.
Also, $\int Q(a)$ = 0 for any $a$.

Let us take a close look at the $*$-multiplication. As discussed
in detail in \cite{Witten:1985cc}, for the multiplication
to be associative, i.e. $(a * a') * a'' = a * (a' * a'')$, we must
interpret the $*$-operation as gluing two half-strings together.
Take two strings $S$, $S'$, whose excitations are described by the
string fields $a$ and $a'$, respectively. Each string is labeled
by a coordinate $0 \leq \sigma \leq \pi$ with the midpoint $\sigma
= \pi/2$. Then the gluing procedure is as follows: The right hand
piece $\pi/2 \leq \sigma \leq \pi$ of the string $S$ and the left
hand piece $0 \leq \sigma \leq \pi/2$ of the string $S'$ are glued
together, and what is left is the string-like object
consisting of the left half of $S$ and the right half of $S'$.
This is the product $S * S'$ in the gluing prescription, and the
resulting string state on $S * S'$ corresponds the string field $a
* a'$. Since $a*a'$ and $a'*a$ are in general thought of as
representing completely different elements, their agreement under
the integration $\int (a*a') = (-1)^{aa'} \int (a' *a)$ (up to a
sing $(-1)^{aa'}$) suggests that the integration procedure still
glues the remaining sides of $S$ and $S'$. If we restate it for a
single string $S*S'$, the left hand piece is sewn to the right
hand piece under the integration.

Using the above definition of the $*$-multiplication and the
integration $\int$, we can write the string field theory action as
follows
\begin{eqnarray}
\label{action1} \mathcal{S}=-\frac{1}{g^2} \int \Big(
\frac{1}{2}\Phi*Q_B \Phi +\frac{1}{3} \Phi*\Phi*\Phi \Big),
\end{eqnarray}
where $g$ is the open string coupling constant, $\Phi$ is the
string field which belongs to the full Hilbert space of the first
quantized open string theory. The algebra is equipped with a
$\mathbf{Z}_2$ grading given by the ghost number, if we define
$\#_\text{gh}$ as an operator that counts the ghost number of its
argument, then we have that: $\#_\text{gh}(\Phi)=1 \, , \,
\#_\text{gh}(Q_B)=1 \, , \, \#_\text{gh}(*)=0 \, , \,
\#_\text{gh}(b)=-1 \, , \, \#_\text{gh}(c)=1$. The action
(\ref{action1}) is invariant under the infinitesimal gauge
transformation $\delta \Phi= Q_B \Lambda +
\Phi*\Lambda-\Lambda*\Phi$, where $\Lambda$ is a gauge parameter
with $\#_\text{gh}(\Lambda) = 0$. In the conformal field theory
(CFT) prescription, the action (\ref{action1}) is evaluated as the
two- and three-point correlation function.

Since the action (\ref{action1}) has been derived quite formally,
it is not suitable for concrete calculations. In particular, $*$
and $\int$ operations have been defined only geometrically as the
gluing procedure. In the next subsection we will argue the methods
of computation based on conformal field theory techniques.

\subsection{The two- and three-string vertex}
The open string worldsheet is parameterized by a complex strip
coordinate $w=\sigma + i \tau$, $\sigma \in [0,\pi]$ or by
$z=-e^{-iw}$ which takes values in the upper half plane (UHP). As
shown in ref. \cite{Schnabl:2005gv}, the gluing conditions
entering into the geometrical definition of the star product
simplify if one uses another coordinate system, $\tilde z = \arctan
z$, in which the upper half plane looks as a semi-infinite
cylinder $C_\pi$ of circumference $\pi$, we have called this new
coordinate system as the sliver frame.

For purposes of computations, the sliver frame seems to be the
most natural one since the conformal field theory in this new
coordinate system remains easy. As in the case of the upper half
plane, we can define general $n$-point correlation functions on
$C_\pi$ which can be readily found in terms of correlation
functions defined on the upper half plane by a conformal mapping,
\begin{eqnarray}
\langle \phi_1(\tilde x_1) \cdots \phi_n(\tilde
x_n)\rangle_{C_\pi}=\langle\tilde\phi_1(\tilde x_1) \cdots
\tilde\phi_n(\tilde x_n)\rangle_{UHP} \; ,
\end{eqnarray}
where the fields $\tilde\phi_i(\tilde x_i)$ are defined as
conformal transformation $\tilde\phi_i(\tilde x_i)= \tan \circ \,
\phi_i(\tilde x_i)$. In general $f \circ \mathcal{\phi}$ denotes a
conformal transformation of a field $\phi$ under a map $f$, for
instance if $\phi$ represents a primary field of dimension $h$,
then $f \circ \mathcal{\phi}$ is defined as $f \circ
\mathcal{\phi}(x)=(f'(x))^h \phi(f(x))$.

The two-string vertex which appears in the string field theory action is
the familiar BPZ inner product of conformal field theory
\footnote{Recall that the BPZ conjugate for the modes of an
holomorphic field $\phi$ of dimension $h$ is given by
bpz$(\phi_n)=(-1)^{n+h}\phi_{-n}$.}. It is defined as a map
$\mathcal{H} \otimes \mathcal{H} \rightarrow \mathbb{R}$
\begin{eqnarray}
\langle\phi_1,\phi_2\rangle = \langle \mathcal{I} \circ \,
\phi_1(0) \phi_2(0) \rangle_{UHP} \; ,
\end{eqnarray}
where $\mathcal{I}: z \rightarrow -1/z$ is the inversion symmetry.
For states defined on the sliver frame $|\tilde\phi_i>$ the
two-string vertex can be written as
\begin{eqnarray}
\langle \tilde\phi_1,\tilde\phi_2\rangle = \langle \mathcal{I}
\circ \, \tilde \phi_1(0) \tilde
\phi_2(0)\rangle_{UHP}=\langle\phi_1(\frac{\pi}{2})
\phi_2(0)\rangle_{C_{\pi}} \, .
\end{eqnarray}
As we can see in this last expression, we evaluate the two-string vertex
at two different points, namely at $\pi/2$ and $0$ on $C_{\pi}$. This must be the case since the inversion symmetry maps the point
at $z=0$ on the upper half plane to the point at infinity,
but the point at infinity is mapped to the point $\pm \pi/2$ on
$C_{\pi}$.

The three-string vertex is a map $\mathcal{H} \otimes
\mathcal{H} \otimes \mathcal{H} \rightarrow \mathbb{R}$, and it is
defined as a correlator on a surface formed by gluing together three
strips representing three open strings. For states defined on the
sliver frame $|\tilde\phi_i>$ the three-string vertex can be written as
\begin{eqnarray}
\langle\tilde\phi_1,\tilde\phi_2,\tilde\phi_3\rangle = \langle
\phi_1(\frac{3 \pi}{4}) \phi_2(\frac{\pi}{4})
\phi_3(-\frac{\pi}{4})\rangle_{C_{   \frac{3 \pi}{2}   }} \, .
\end{eqnarray}
Here the correlator is taken on a semi-infinite cylinder $C_{
\frac{3 \pi}{2}}$ of circumference $ 3 \pi/2$. Also, this
correlator can be evaluated on the semi-infinite cylinder $C_\pi$
of circumference $\pi$. We only need to perform a simple conformal
map (scaling) $s:\tilde z \rightarrow \frac{2}{3} \tilde z$ which
brings the region $C_{ \frac{3 \pi}{2}}$ to $C_\pi$, and the
correlator is given by
\begin{eqnarray}
\langle\tilde\phi_1,\tilde\phi_2,\tilde\phi_3\rangle = \langle s
\circ \, \phi_1(\frac{3 \pi}{4}) s \circ \, \phi_2(\frac{\pi}{4})
s \circ \, \phi_3(-\frac{\pi}{4})\rangle_{C_\pi} \, .
\end{eqnarray}
Note that the scaling transformation $s$ is implemented by $U_3 =
(2/3)^{\mathcal{L}_0}$, where $\mathcal{L}_0$ is the zero mode of
the worldsheet energy momentum tensor $T_{\tilde z \tilde
z}(\tilde z)$ in the $\tilde z$ coordinate,
\begin{eqnarray}
\mathcal{L}_0=\oint \frac{d \tilde z}{2 \pi i} \tilde z T_{\tilde
z \tilde z}(\tilde z) \; .
\end{eqnarray}
By a conformal transformation it can be expressed as
\begin{eqnarray}
\mathcal{L}_0= \oint \frac{d z}{2 \pi i} (1+z^2) \arctan z T_{ z
z}(z)=L_0+\sum_{k=1}^{\infty} \frac{2(-1)^{k+1}}{4k^2-1}L_{2k} \;
,
\end{eqnarray}
where the $L_n$'s are the ordinary Virasoro generators (with zero
central charge) of the full (matter plus ghost) conformal field
theory.

\subsection{Correlation functions}
In this subsection we list correlation functions
evaluated on the semi-infinite cylinder $C_\pi$. As already
mentioned, the relation between correlation functions evaluated on
the upper half plane and those evaluated on the semi-infinite
cylinder is given by conformal transformation.

Employing the definition of the conformal transformation $\tilde
c(x)=\cos^2(x) c(\tan x)$ of the $c$ ghost and its anticommutation
relations with the operators $Q_B$, $\mathcal{B}_0$ and $B_1$,
\footnote{The operators $\mathcal{B}_0$ and
$B_1\equiv\mathcal{B}_{-1}$ are modes of the $b$ ghost which are
defined on the semi-infinite cylinder coordinate as
$\mathcal{B}_{n}=\oint \frac{dz}{2 \pi i}(1+z^2) (\arctan
z)^{n+1}b(z)$.}
\begin{align}
\{Q_B,\tilde c(z)\}&= \tilde c(z) \partial \tilde c(z) \, , \\
\{\mathcal{B}_0,\tilde c(z)\}&=z \, , \\
\{B_1,\tilde c(z)\}&=1 \, ,
\end{align}
we obtain the following basic correlation functions,
\begin{align}
\label{a0} \langle \tilde{c}(x)\tilde{c}(y)\tilde{c}(z) \rangle
&=\sin(x -y)\sin(x - z)\sin(y - z) \, , \\
\label{a1} \langle \tilde c(x) Q_B \tilde c(y) \rangle &= - \sin(x-y)^2 \, , \\
\label{a2}\langle \tilde{c}(x)\mathcal{B}_0
\tilde{c}(y)\tilde{c}(z) \tilde{c}(w)\rangle &= y \langle
\tilde{c}(x)\tilde{c}(z)\tilde{c}(w)\rangle-z \langle
\tilde{c}(x)\tilde{c}(y)\tilde{c}(w)\rangle +w \langle
\tilde{c}(x)\tilde{c}(y)\tilde{c}(z) \rangle \, , \\
\label{a3}\langle \tilde{c}(x)\tilde{c}(y)\mathcal{B}_0
\tilde{c}(z) \tilde{c}(w)\rangle &= z \langle
\tilde{c}(x)\tilde{c}(y)\tilde{c}(w)\rangle -w \langle
\tilde{c}(x)\tilde{c}(y)\tilde{c}(z) \rangle \, , \\
\label{a4}\langle \tilde{c}(x)B_1 \tilde{c}(y)\tilde{c}(z)
\tilde{c}(w)\rangle &= \langle
\tilde{c}(x)\tilde{c}(z)\tilde{c}(w)\rangle-\langle
\tilde{c}(x)\tilde{c}(y)\tilde{c}(w)\rangle + \langle
\tilde{c}(x)\tilde{c}(y)\tilde{c}(z) \rangle \, , \\
\label{a5}\langle \tilde{c}(x)\tilde{c}(y)B_1 \tilde{c}(z)
\tilde{c}(w)\rangle &= \langle
\tilde{c}(x)\tilde{c}(y)\tilde{c}(w)\rangle - \langle
\tilde{c}(x)\tilde{c}(y)\tilde{c}(z) \rangle \, .
\end{align}

To compute correlation functions involved in the evaluation of the
string field theory action, the following contour integrals will
be very useful,
\begin{align}
\label{sa}\sigma(a)&\equiv \oint \frac{dz}{2 \pi i} z^{a} \sin(2z) \nonumber \\
&= \frac{\theta(-a-2)}{\Gamma(-a)} ((-1)^{a}+1)
(-1)^{\frac{2-a}{2}} 2^{-a-2} \, , \\
\label{ca}\varsigma(a)&\equiv \oint \frac{dz}{2 \pi i} z^{a} \cos(2z) \nonumber \\
&= \frac{\theta(-a-1)}{\Gamma(-a)} ((-1)^{a}-1)
(-1)^{\frac{1-a}{2}} 2^{-a-2} \, ,
\end{align}
\begin{align}
\label{f}
\mathcal{F}(a_1,a_2,a_3,\alpha_1,\beta_1,\alpha_2,\beta_2,\alpha_3,\beta_3)\equiv
\oint \frac{dx_1 dx_2 dx_3}{(2 \pi i)^3}
x_1^{a_1}x_2^{a_2}x_3^{a_3} \langle \tilde c(\alpha_1 x_1
+\beta_1) \tilde c(\alpha_2 x_2 +\beta_2)\tilde c(\alpha_3 x_3
+\beta_3) \rangle \nonumber \;\;\;\;\;\;\;\;\;\;\\
= \frac{1}{ \alpha_1^{a_1+1}\alpha_2^{a_2+1}\alpha_3^{a_3+1}}
\Big[
\;\;\;\;\;\;\;\;\;\;\;\;\;\;\;\;\;\;\;\;\;\;\;\;\;\;\;\;\;\;\;\;\;\;\;\;\;\;\;\;\;\;\;\;\;\;\;\;\;\;\;\;\;\;
 \;\;\;\;\;\;\;\;\;\;\;\;\;\;\;\;\;\;\;\;\;\;\;\;   \nonumber\\
 \delta_{a_3,-1}\frac{ \big(\sigma(a_1)\sigma(a_2)+\varsigma(a_1)\varsigma(a_2)\big) \sin(2(\beta_1-\beta_2))
 +
 \big(\sigma(a_1)\varsigma(a_2)-\varsigma(a_1)\sigma(a_2)\big) \cos(2(\beta_1-\beta_2))}{4} \nonumber \;\;\;\;\;\;\;\;\;\;\;\;\;\;\;\;\;\;\;\;\\
 + \delta_{a_2,-1}\frac{ \big(\varsigma(a_1)\sigma(a_3)-\sigma(a_1)\varsigma(a_3)\big) \cos(2(\beta_1-\beta_3))
 -
 \big(\varsigma(a_1)\varsigma(a_3)+\sigma(a_1)\sigma(a_3)\big) \sin(2(\beta_1-\beta_3))}{4} \nonumber \;\;\;\;\;\;\;\;\;\;\;\;\;\;\;\;\;\;\\
 + \delta_{a_1,-1}\frac{ \big(\varsigma(a_2)\mathcal{\varsigma}(a_3)+\sigma(a_2)\sigma(a_3)\big) \sin(2(\beta_2-\beta_3))
 +
 \big(\sigma(a_2)\varsigma(a_3)-\varsigma(a_2)\sigma(a_3)\big)
 \cos(2(\beta_2-\beta_3))}{4}
 \Big], \;\;\;\;\;\;\;\;\;\;\;\;\;\;\; \nonumber \\
\end{align}
where $\theta(n)$ is the unit step (Heaviside) function which is
defined as
\begin{align}
\theta(n) =
\begin{cases}
  0,  & \mbox{if }n < 0 \\
  1, & \mbox{if }n \geq 0 \, .
\end{cases}
\end{align}

Let us list a few non-trivial correlation functions which involve
operators frequently used in the $\mathcal{L}_0$ basis, namely,
$\hat{\mathcal{L}}^{n}$ ($\hat{\mathcal{L}}\equiv {\cal L}_0+{\cal
L}_0^\dag$), $\hat{\mathcal{B}}$ ($\hat{\mathcal{B}}\equiv {\cal
B}_0+{\cal B}_0^\dag$), $U_r=\big(\frac{2}{r}\big)^{{\cal L}_0}$
and the  $\tilde c(z)$ ghost
\begin{align}
\label{correla1}&\langle \text{bpz}(\tilde{c}_{p_1})
\hat{\mathcal{L}}^{n_1} U^\dag_{r} U_{r}
\tilde{c}(x)\tilde{c}(y) \rangle = \nonumber \\
&= \oint \frac{dz_1 dx_1}{(2 \pi i)^{2}}\frac{(-2)^{n_1} n_1! \,
x_1^{p_1-2}}{(z_1-2)^{n_1+1}} \big(\frac{2}{r}\big)^{-p_1+n_1-2}
\big(\frac{2}{z_1}\big)^{-p_1-2} \langle
\tilde{c}(x_1+\frac{\pi}{2}) \tilde{c}(\frac{4}{z_1
r}x)\tilde{c}(\frac{4}{z_1 r}y) \rangle \, ,
\end{align}
\begin{align}
\label{correla2}&\langle \text{bpz}(\tilde{c}_{p_1})
\hat{\mathcal{L}}^{n_1} \hat{\mathcal{B}} U^\dag_{r} U_{r}
\tilde{c}(x)\tilde{c}(y)\tilde{c}(z) \rangle = \nonumber \\
&= - \delta_{p_1,0} \oint \frac{dz_1}{ 2 \pi i} \frac{(-2)^{n_1}
n_1!}{(z_1-2)^{n_1+1}} \big(\frac{2}{r}\big)^{-p_1+n_1-2}
\big(\frac{2}{z_1}\big)^{-p_1-2} \langle \tilde{c}(\frac{4}{z_1
r}x)\tilde{c}(\frac{4}{z_1 r}y)\tilde{c}(\frac{4}{z_1 r}z) \rangle
\nonumber \\
&+  \oint \frac{dz_1 dx_1}{(2 \pi i)^{2}}\frac{(-2)^{n_1} n_1! \,
x_1^{p_1-2}}{(z_1-2)^{n_1+1}} \big(\frac{2}{r}\big)^{-p_1+n_1-2}
\big(\frac{2}{z_1}\big)^{-p_1-2} \langle
\tilde{c}(x_1+\frac{\pi}{2}) \mathcal{B}_0 \tilde{c}(\frac{4}{z_1
r}x)\tilde{c}(\frac{4}{z_1 r}y)\tilde{c}(\frac{4}{z_1 r}z) \rangle
\, ,
\end{align}
\begin{align}
\label{correla3}&\langle \text{bpz}(\tilde{c}_{p_1})
\text{bpz}(\tilde{c}_{p_2}) \hat{\mathcal{L}}^{n_1}
\hat{\mathcal{B}} U^\dag_{r} U_{r}
\tilde{c}(x)\tilde{c}(y) \rangle = \nonumber \\
&= - \delta_{p_2,0}\oint \frac{dz_1 dx_1}{(2 \pi
i)^{2}}\frac{(-2)^{n_1} n_1! \, x_1^{p_1-2}}{(z_1-2)^{n_1+1}}
\big(\frac{2}{r}\big)^{-p_1-p_2+n_1-1}
\big(\frac{2}{z_1}\big)^{-p_1-p_2-1} \langle
\tilde{c}(x_1+\frac{\pi}{2}) \tilde{c}(\frac{4}{z_1
r}x)\tilde{c}(\frac{4}{z_1 r}y) \rangle \nonumber \\
& +\delta_{p_1,0}\oint \frac{dz_1 dx_2}{(2 \pi
i)^{2}}\frac{(-2)^{n_1} n_1! \, x_2^{p_2-2}}{(z_1-2)^{n_1+1}}
\big(\frac{2}{r}\big)^{-p_1-p_2+n_1-1}
\big(\frac{2}{z_1}\big)^{-p_1-p_2-1} \langle
\tilde{c}(x_2+\frac{\pi}{2}) \tilde{c}(\frac{4}{z_1
r}x)\tilde{c}(\frac{4}{z_1 r}y) \rangle \nonumber \\
& +\oint \frac{dz_1 dx_1 dx_2}{(2 \pi i)^{3}}\frac{(-2)^{n_1} n_1!
\, x_1^{p_1-2} x_2^{p_2-2}}{(z_1-2)^{n_1+1}}
\big(\frac{2}{r}\big)^{-p_1-p_2+n_1-1}
\big(\frac{2}{z_1}\big)^{-p_1-p_2-1} \times \nonumber \\
 &\;\;\;\;\;\;\;\;\;\;\;\;\;\;\;\;\;\;\;\;\;\;\;\;\;\;\;\;
 \;\;\;\;\;\;\;\;\;\;\;\;\;\;\;\;\;\;\;\;\;\;\;
 \;\;\;\;\;\;\;\;\;\;\;\;\;\;\;\;\;\; \times \langle \tilde{c}(x_1+\frac{\pi}{2})
\tilde{c}(x_2+\frac{\pi}{2}) \mathcal{B}_0 \tilde{c}(\frac{4}{z_1
r}x)\tilde{c}(\frac{4}{z_1 r}y) \rangle \, ,
\end{align}
where the ``\emph{bpz}'' acting on the modes of the $\tilde c(z)$
ghost stands for the usual BPZ conjugation which in the ${\cal
L}_0$ basis is defined as follows
\begin{align}
\text{bpz}(\tilde \phi_n) &= \oint \frac{d  z}{2 \pi i} z^{n+h-1}
\tilde \phi ( z + \frac{\pi}{2}) \, ,
\end{align}
for any primary field $\tilde \phi(z)$ with weight $h$. The action
of the BPZ conjugation on the modes of $\tilde \phi(z)$ satisfies
the following useful property
\begin{align}
U_r^{\dag -1} \text{bpz}(\tilde \phi_n)
U_r^{\dag}=\big(\frac{2}{r}\big)^{-n} \text{bpz}(\tilde \phi_n) \,
.
\end{align}
Correlation functions which involve only modes of the $\tilde
c(z)$ ghost can be expressed in terms of the contour integral
(\ref{f}) as follows
\begin{align}
\label{co1}\langle \tilde{c}_p \tilde{c}_q \tilde{c}_r \rangle &=
\oint \frac{dx dy dz}{(2 \pi i)^{3}} x^{p-2} y^{q-2} z^{r-2}
\langle \tilde{c}(x)\tilde{c}(y)\tilde{c}(z) \rangle \nonumber \\
&= \mathcal{F}(p-2,q-2,r-2,1,0,1,0,1,0) \, ,
\\
\label{corre2}\langle \text{bpz}(\tilde{c}_p) \tilde{c}_q
\tilde{c}_r \rangle &= \oint \frac{dx dy dz}{(2 \pi i)^{3}}
x^{p-2} y^{q-2} z^{r-2}
\langle \tilde{c}(x+\frac{\pi}{2})\tilde{c}(y)\tilde{c}(z) \rangle \nonumber \\
&= \mathcal{F}(p-2,q-2,r-2,1,\frac{\pi}{2},1,0,1,0)  \, ,
\\
\label{corre3}\langle \text{bpz}(\tilde{c}_p)
\text{bpz}(\tilde{c}_q) \tilde{c}_r \rangle &= \oint \frac{dx dy
dz}{(2 \pi i)^{3}} x^{p-2} y^{q-2} z^{r-2}
\langle \tilde{c}(x+\frac{\pi}{2})\tilde{c}(y+\frac{\pi}{2})\tilde{c}(z) \rangle \nonumber \\
&= \mathcal{F}(p-2,q-2,r-2,1,\frac{\pi}{2},1,\frac{\pi}{2},1,0) \,
.
\end{align}

To evaluate correlators involving modes of the $\tilde c(z)$ ghost
and insertions of operators $\hat{\mathcal{L}}^{n}$,
$\hat{\mathcal{B}}$, we can use the basic correlators
(\ref{a2})-(\ref{a5}) and the definition of $\hat{\mathcal{L}}^{n}
\equiv (-2)^n n! \oint \frac{dz}{2 \pi i} \frac{1}{(z-2)^{n+1}}
U^\dag_z U_z$. For instance, as a pedagogical illustration let us
compute a correlator involving a $\hat{\mathcal{L}}^{n}$ insertion,
\begin{align}
\label{corre5}\langle
\text{bpz}(\tilde{c}_p)(\mathcal{L}_0+\mathcal{L}^\dag_0)^{n}
\tilde{c}_q \tilde{c}_r \rangle &=(-2)^n n! \oint \frac{dz_1}{2
\pi i} \frac{1}{(z_1-2)^{n+1}} \langle \text{bpz}(\tilde{c}_p)
U^\dag_{z_1} U_{z_1} \tilde{c}_q \tilde{c}_r \rangle \nonumber\\
&=(-2)^n n! \oint \frac{dz_1}{2 \pi i}
\frac{(\frac{2}{z_1})^{-p-q-r}}{(z_1-2)^{n+1}} \langle
\text{bpz}(\tilde{c}_p) \tilde{c}_q \tilde{c}_r \rangle \nonumber
\\
&=(-1)^n n! {p+q+r \choose n}
\mathcal{F}(p-2,q-2,r-2,1,\frac{\pi}{2},1,0,1,0) \, ,
\end{align}
where we have used the following useful contour integral $\;\oint
\frac{dz}{2 \pi i} \frac{z^{m}}{(z-a)^{n+1}} = {m \choose n}
a^{m-n}$.

Correlators involving the $*$-product can be computed using the
results of this subsection. For instance, let us compute the
correlator $\langle 0|\text{bpz}(\tilde c_{p_1})
\hat{\mathcal{L}}^{n_1} ,\hat{\mathcal{L}}^{n_2}  \tilde c_{p_2}
|0 \rangle * \hat{\mathcal{L}}^{n_3}  \tilde c_{p_3}|0 \rangle$,
\begin{align}
&\label{corre7} \langle 0|\text{bpz}(\tilde c_{p_1})
\hat{\mathcal{L}}^{n_1} ,\hat{\mathcal{L}}^{n_2}  \tilde c_{p_2}
|0 \rangle * \hat{\mathcal{L}}^{n_3}  \tilde c_{p_3}|0 \rangle =
\nonumber \\
&= \frac{(-2)^{n_2+n_3} n_2! n_3! }{(2 \pi i)^{4}} \oint
\frac{dz_2 dz_3 dx_2 dx_3 \,
x_2^{p_2-2}x_3^{p_3-2}}{(z_2-2)^{n_2+1}(z_3-2)^{n_3+1}} \langle
0|\text{bpz}(\tilde c_{p_1}) \hat{\mathcal{L}}^{n_1}, U^\dag_{z_2}
U_{z_2} \tilde c(x_2)|0 \rangle * U^\dag_{z_3} U_{z_3} \tilde
c(x_3) |0 \rangle
\nonumber \\
&= \frac{(-2)^{n_2+n_3} n_2! n_3! }{(2 \pi i)^{4}} \oint
\frac{dz_2 dz_3 dx_2 dx_3 \,
x_2^{p_2-2}x_3^{p_3-2}}{(z_2-2)^{n_2+1}(z_3-2)^{n_3+1}} \times
\nonumber
\\ &\times \langle \text{bpz}(\tilde c_{p_1}) \hat{\mathcal{L}}^{n_1}
U^\dag_{r} U_{r} \tilde c(x_2+\frac{\pi}{4}(z_3-1))
\tilde c(x_3-\frac{\pi}{4}(z_2-1)) \rangle \nonumber \\
&=\frac{(-1)^{n_1+n_2+n_3} 2^{2 n_1+n_2+n_3-2p_1-4} n_1!n_2! n_3!
}{(2 \pi i)^{3}} \oint \frac{dz_1dz_2 dz_3 \,
z_1^{p_1+2}r^{p_1+2-n_1}}{(z_1-2)^{n_1+1}(z_2-2)^{n_2+1}(z_3-2)^{n_3+1}}
\times \nonumber
\\ &\times \mathcal{F}(p_1-2,p_2-2,p_3-2,1,\frac{\pi}{2},\frac{4}{z_1r},\frac{\pi(z_3-1)}{z_1r},
\frac{4}{z_1r},\frac{\pi(1-z_2)}{z_1r}) \, ,
\end{align}
where we have defined $r\equiv z_2+z_3-1$.

\section{The tachyon potential}
To compute the effective tachyon potential in a particular gauge, it is
necessary to specify which fields are being integrated out.
Usually, when we compute the effective tachyon potential in the
Siegel gauge, the fields which are integrated out correspond to
the perturbative Fock space of states with mass greater than the
tachyon mass \cite{Moeller:2000xv,Taylor:2002fy}.

In this section, in order to find the effective tachyon potential
in the sliver frame, instead of integrating out fields in the
state space mentioned in the previous paragraph, we are going to
integrate out fields with $\mathcal{L}_0$ eigenvalue greater than
the $\mathcal{L}_0$ eigenvalue of the tachyon state $\tilde c_1
|0\rangle$. This means that the effective tachyon potential we
compute is different from the old effective tachyon potential
computed in the Siegel gauge. As we already commented in the
introduction, we choose the state $\tilde c_1 |0\rangle$ as the
tachyonic state since it is the most natural one from the
perspective of Schnabl's coordinates (the sliver frame). Choosing
insertions on other wedge states does not seem to be natural,
except for insertions over the sliver or the identity $\tilde c_1
|\infty\rangle$, $\tilde c_1 |\mathcal{I}\rangle$, nevertheless
both of these options are singular
\cite{Kluson:2002ex,Kishimoto:2001de,Takahashi:2002ez,Rastelli:2001jb}.

\subsection{The effective tachyon potential in the Schnabl gauge}
As in the case of the Siegel gauge, in the Schnabl gauge we could
perform an analysis of the tachyon potential by performing
computations in the $\mathcal{L}_0$ level truncation. We are going
to define the level of a state as the eigenvalue of the operator
$N=\mathcal{L}_0+1$. This definition is adjusted so that the zero
momentum tachyon $\tilde c_1 |0\rangle$ is at level zero.

Having defined the level number of states contained in the level
expansion of the string field, level of each term in the action is
also defined to be the sum of the levels of the fields involved.
For instance, if states $\tilde \phi_1$, $\tilde \phi_2$, $\tilde
\phi_3$ have level $n_1$, $n_2$, $n_3$ respectively, we assign
level $n_1+n_2+n_3$ to the interaction term
$\langle\tilde\phi_1,\tilde\phi_2,\tilde\phi_3\rangle$. When we
say level $(m,n)$, we mean that the string field includes all
terms with level $\leq m$ while the action includes all terms with
level $\leq n$.

In this paper, we want to study questions related to the
appearance of a stable vacuum in the theory when the tachyon and
other scalar fields acquire nonzero expectation values. Because
all the questions we will address involve Lorentz-invariant
phenomena, we can restrict attention to scalar fields in the
string field expansion. We write the string field expansion in
terms of scalar fields as
\begin{eqnarray}
\Psi= \sum_{i=0}^{\infty} x_{i} | \psi^i \rangle \, ,
\end{eqnarray}
where in the $\mathcal{L}_0$ level expansion the state $| \psi^i
\rangle$ is built by applying the modes of the $\tilde c(z)$ ghost
and the operators $({\cal L}_0+{\cal L}_0^\dag)^n$, ${\cal
B}_0+{\cal B}_0^\dag$ on the $SL(2,\mathbb{R})$ invariant vacuum
$|0\rangle$. The first term in the expansion is given by the
zero-momentum tachyon $| \psi^0 \rangle =\tilde c_1 |0\rangle $.
We will restrict our attention to an even-twist and ghost-number
one string field $\Psi$ satisfying the Schnabl gauge
$\mathcal{B}_0 \Psi=0$. Choosing a particular gauge prevents the
inclusion of `almost' flat directions which would have correspond
to gauge degrees of freedom in the potential (`almost' is in
quotation marks, since level truncation destroys gauge symmetry).
The tachyon potential we want to evaluate is defined as
\begin{eqnarray}
\label{norpoten1} V=2 \pi^2 \Big[\frac{1}{2} \langle \Psi,Q_B \Psi
\rangle +\frac{1}{3} \langle \Psi,\Psi*\Psi \rangle  \Big] \, .
\end{eqnarray}
The effective tachyon potential can be determined by starting with
the complete set of terms in the potential truncated at some level
$(m,n)$, fixing a value for $x_0$, solving for all coefficients
$x_i$, $i \geq 1$, and plugging them back into the potential to rewrite
it as a function of $x_0$.

In order to explain the procedure for finding the effective
tachyon potential, let us first set all components of
the string field $\Psi$ to zero except for the first coefficient
$x_0$. This state will be said to be of level zero. Thus, we take
\begin{eqnarray}
\label{leve0} \Psi= x_0 \tilde c_1  |0\rangle \, .
\end{eqnarray}
Plugging (\ref{leve0}) into the definition (\ref{norpoten1}), we
get the zeroth approximation to the tachyon potential,
\begin{eqnarray}
\label{potent0} V^{(0,0)} =2 \pi^2 \Big[ -\frac{x_0^2}{2}+\frac{27
\sqrt{3} x_0^3}{64} \Big] \, .
\end{eqnarray}

To compute corrections to this result, we need to include higher
level fields in our analysis. The analysis can be simplified by
noting that the potential (\ref{norpoten1}) has a twist symmetry
under which all coefficients of odd-twist states change sign,
whereas coefficients of even-twist states remain unchanged.
Therefore coefficients of odd-twist states at levels above $\tilde
c_1|0\rangle$ must always appear in the action in pairs, and they
trivially satisfy the equations of motion if set to zero. Thus,
we look for $\Psi$ containing only
even-twist states.

Taking into account the considerations above, at the next level we
find that the string field is given by
\begin{eqnarray}
\label{leve1} \Psi= x_0 \tilde c_1  |0\rangle - 2 x_1 ({\cal
L}_0+{\cal L}_0^\dag) \tilde c_1 |0\rangle - 2 x_1 ({\cal
B}_0+{\cal B}_0^\dag) \tilde c_0 \tilde c_1 |0\rangle \, ,
\end{eqnarray}
where the coefficients of the expansion were chosen so that $\Psi$
satisfies the Schnabl gauge, $\mathcal{B}_0 \Psi=0$. Substituting
this level expansion of the string field (\ref{leve1}) into
(\ref{norpoten1}) we get the (1,3) level approximation to the
potential
\begin{eqnarray}
\label{potent1} V^{(1,3)} =2 \pi^2 \Big[ -\frac{x_0^2}{2}+\frac{27
\sqrt{3} x_0^3}{64}+\big(\frac{27}{8} \sqrt{3} -\frac{9}{8} \pi
\big)x_0^2 x_1+ \big(\frac{9}{2} \sqrt{3} -3 \pi +\frac{2
\pi^2}{\sqrt{3}} \big)x_0 x_1^2 \Big] \, .
\end{eqnarray}

Since the effective tachyon potential depends on the single
variable $x_0$ which corresponds to the tachyon coefficient, we
are going to integrate out the variable $x_1$. Using the partial
derivative of the potential, $\partial_{x_1}V^{(1,3)}=0$, we can
write the variable $x_1$ in terms of $x_0$
\begin{eqnarray}
\label{x1} x_1=\frac{27 \pi-81 \sqrt{3}}{216 \sqrt{3}-144 \pi +32
\sqrt{3} \pi ^2}\, x_0 \, .
\end{eqnarray}
By plugging back (\ref{x1}) into the potential (\ref{potent1}) to
rewrite it as a function of the single variable $x_0$, we obtain
the effective potential
\begin{eqnarray}
\label{potefec1} V^{(1,3)}_{eff} =2 \pi^2 \Big[
-\frac{x_0^2}{2}+\frac{486 \sqrt{3} \pi -2187+405 \pi ^2}{3456
\sqrt{3}-2304 \pi +512 \sqrt{3} \pi ^2} \, x_0^3 \Big] \, .
\end{eqnarray}

Extending our analysis to the next level, we are going to use the
string field $\Psi$ satisfying the Schnabl gauge ${\cal B}_0
\Psi=0$ expanded up to level two states,
\begin{eqnarray}
\Psi&=& x_0 \tilde c_1  |0\rangle - 2 x_1 ({\cal L}_0+{\cal
L}_0^\dag) \tilde c_1 |0\rangle - 2 x_1 ({\cal B}_0+{\cal
B}_0^\dag) \tilde c_0 \tilde c_1 |0\rangle + x_2 \tilde c_{-1}|0\rangle \nonumber \\
\label{leve2} &-&x_3({\cal L}_0+{\cal L}_0^\dag)^2 \tilde
c_1|0\rangle-2 x_3 ({\cal L}_0+{\cal L}_0^\dag) ({\cal B}_0+{\cal
B}_0^\dag) \tilde c_0 \tilde c_1 |0\rangle \, .
\end{eqnarray}
To obtain the level (2,6) potential, we plug the string field
(\ref{leve2}) into the definition (\ref{norpoten1}). By using some
correlation functions derived in the previous section, we arrive
to the following potential
\begin{eqnarray}
\label{potent2} V^{(2,6)}&=&2\pi
^2\Big[-\frac{x_0^2}{2}\;+\;\frac{27 \sqrt{3}
x_0^3}{64}\;+\;\big(\frac{27}{8} \sqrt{3}-\frac{9}{8}
\pi \big)x_0^2 x_1\;+\;\big(\frac{9}{2} \sqrt{3} -3 \pi +\frac{2 \pi ^2 }{\sqrt{3}}\big)x_0 x_1^2 \nonumber \\
&+&x_0 x_2-\frac{3}{4} \sqrt{3} x_0^2 x_2-\frac{16 \pi ^2 x_1^2
x_2}{9 \sqrt{3}}-x_2^2\;+\;\frac{x_0
x_2^2}{\sqrt{3}}\;+\;\big(\frac{8 \pi}{27}-\frac{8}{3 \sqrt{3}}\big)x_1 x_2^2-2 x_0 x_3 \nonumber \\
&+&\big(\frac{3\pi}{4} -\frac{9}{8} \sqrt{3} +\frac{5 \pi ^2 }{8
\sqrt{3}}\big)x_0^2x_3-\frac{8 \pi ^4 x_1^2 x_3}{81
\sqrt{3}}-\frac{7 \pi ^2 x_0 x_2 x_3}{9 \sqrt{3}}+
\big(\frac{8}{243} \pi ^3 -\frac{8 \pi ^2}{3 \sqrt{3}}\big)x_1
x_2 x_3 \nonumber \\
&+&\big(\frac{16 \pi ^2 }{81 \sqrt{3}}-\frac{8 }{3
\sqrt{3}}+\frac{16\pi}{27}\big)x_2^2 x_3+\big(\frac{2 \pi ^5
}{2187}-\frac{98 \pi ^4}{243 \sqrt{3}}\big) x_1 x_3^2 + \frac{17
\pi ^4 x_0 x_3^2}{324 \sqrt{3}}\nonumber \\
&+&\big(\frac{16\pi ^3}{243}-\frac{8 \pi ^2}{3 \sqrt{3}}-\frac{16
\pi ^4}{243 \sqrt{3}}\big)x_2 x_3^2+\big(\frac{4 \pi
^5}{2187}-\frac{98 \pi ^4}{243 \sqrt{3}}-\frac{28 \pi ^6}{6561
\sqrt{3}}\big)x_3^3 \; \Big].
\end{eqnarray}

As we can see, starting at level (2,6), coefficients other than
the tachyon coefficient $x_0$ are no longer quadratic, therefore
we cannot exactly integrate out all these non-tachyonic
coefficients ($x_i$, $i \geq 1$). Therefore, we are forced to use
numerical methods to study the effective tachyon potential. We
have used Newton's method to find the zeros of the partial
derivatives of the potential. For a fixed value of the tachyon
coefficient $x_0$, there are in general many solutions of the
equations for the remaining coefficients $x_i$, $i \geq 1$, which
correspond to different branches of the effective potential. We
are interested in the branch connecting the perturbative with the
nonperturbative vacuum and having a minimum value which agrees
with the one predicted from Sen's first conjecture.

Applying the numerical approach described above, we have
integrated out the variables $x_1$, $x_2$ and $x_3$ appearing in
the potential (\ref{potent2}). At this level, we found that the
shape of the effective tachyon potential which connects the
perturbative with the nonperturbative vacuum is given by the graph
shown in figure \ref{fig:tach2}. For reference we have plotted the
effective tachyon potential up to level (3,9). The minimum value
of the level (2,6) effective potential $V^{(2,6)}_{eff}$ occurs at
$x_{0,min}=0.7023612173$, and its depth gets the value of
$-1.0466220796$ which is $4.66 \%$ greater than the conjectured
value (\ref{conjec1}). At this level, we have noted that our
algorithm becomes unstable for values of the tachyon coefficient
between $0<x_0<0.47$, this may indicate that the branch which
contain the perturbative with the nonperturbative vacuum meets one
or more other branches which play the role of attractors. In fact,
we have found that there is a new branch which meets the physical
branch\footnote{We refer to the physical branch as the branch
which connects the perturbative with the nonperturbative vacuum.}
at $x_0 \approx 0.47$. This new branch contains the extremal
points $x_0=0.2205432494$, $x_1=-0.0150554001$,
$x_2=-0.2576803891$, $x_3=0.1491283766$, which are solutions to
the equations coming from the partial derivatives of the potential
(\ref{potent2}). We have excluded this new branch (generated by
these points) since it does not contain the nonperturbative
vacuum.
\begin{figure}[ht]
\begin{center}
\includegraphics[width=5.88in,height=95mm]{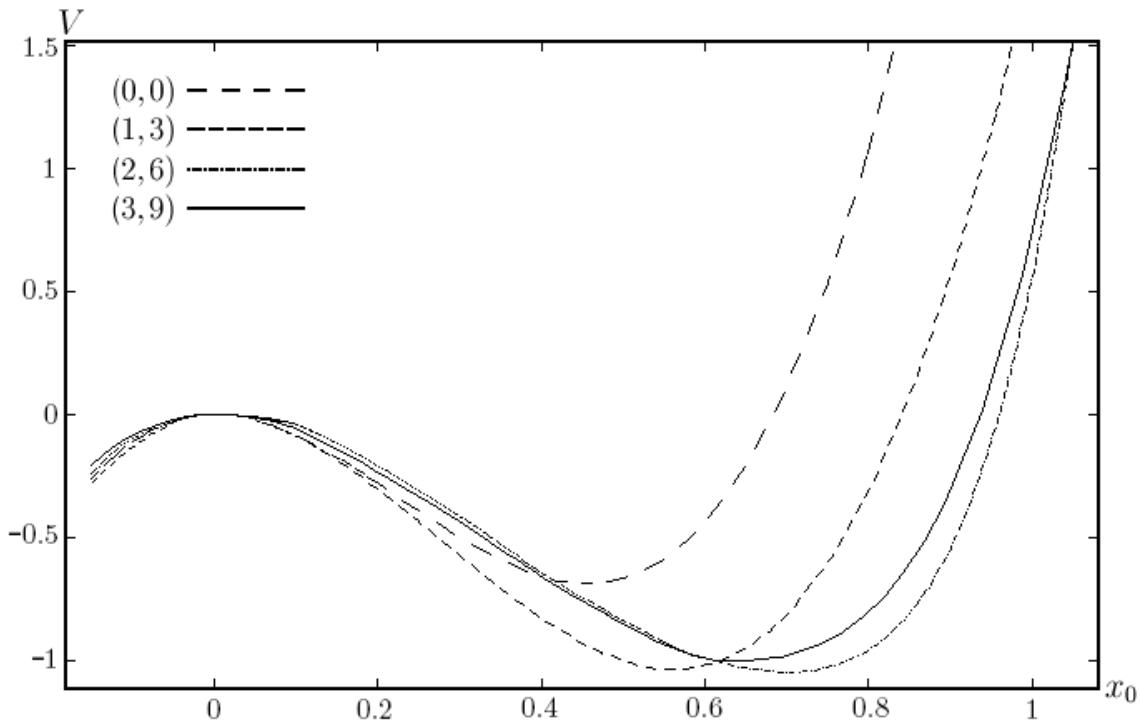}
\end{center}
\caption{Effective tachyon potential at different levels.}
\label{fig:tach2}
\end{figure}

We could continue to perform higher level computations, since these
computations follow the same procedures shown above, but at this point
we only want to comment about the results. Higher level
computations reveal that the effective tachyon potential has the
profile found in the lower level cases. However, at higher levels
the algorithm used to compute the effective tachyon potential
fails to converge outside the region $-0.014<x_0<0.701$. This
result indicates that the effective tachyon potential has branch
points near $x_0\approx -0.014$ and $x_0 \approx 0.701$, where the
nontrivial vacuum appears at $x_0=0.636$. The locations of these
branch points appear to converge under $\mathcal{L}_0$ level
truncation to fixed values.

While the perturbative and nonperturbative vacua both lie on the
effective potential curve between these branch points, the
existence of these branch points prompts us to ask what cubic
string field theory can say about the effective tachyon potential
beyond these branch points. In particular, an issue of some
interest is how the effective tachyon potential behaves for large
negative values of $x_0$. A possible physical reason why our
algorithm fails to converge outside the region $-0.014<x_0<0.701$
might be analogous to the case of the effective potential found in
the usual $L_0$ level expansion, where the the existence of these
branch points is related to the validity of the Siegel gauge. It
is then possible that in the $\mathcal{L}_0$ level expansion the
singularities previously found in the effective tachyon potential
are gauge artifacts arising from the boundary of the region of
validity of Schnabl gauge. We leave the analysis of this
possiblity for further research.

Up to the level that we have explored with our computations, it is
worth remarking that the depth of the effective potential is
converging to the conjectured value (\ref{conjec1}). For instance,
at level (5,15) the minimum value of the effective potential
occurs at $x_{0,min}=0.6368018630$, and its depth takes on the
value $-0.9993346627$ which is $99.93\%$ of the conjectured value.

Another remark is that at the minimum of
the effective tachyon potential, the tachyon coefficient $x_0$ is
approaching the analytical value of $2/\pi$, which interestingly
is the same value of the tachyon coefficient in Schnabl's
analytical solution when expanded in the $\mathcal{L}_0$ basis. In
the next subsection we will give additional comments on this
important observation.

To conclude this subsection, let us note that if we go to high
enough level, the depth of the effective tachyon potential should
diverge since the minimum of the effective potential should
correspond to the value of the action evaluated on Schnabl's
solution truncated at some finite $\mathcal{L}_0$ level. In order
to regularize the value of the minimum of the effective potential,
the use of Pad\'{e} resummation techniques would eventually be
needed \cite{Arroyo:2009ec}.

\subsection{The stable vacuum and Schnabl's solution}
In this subsection we want to address the connection between the
configuration found by extremizing the tachyon potential and
Schnabl's analytic solution.

In order to compare the analytic solution with the one obtained by
the methods shown in the previous subsection, let us write Schnabl's
analytic solution up to level-two states
\begin{eqnarray}
\Psi_{analytic}&=& \frac{2}{\pi} \tilde c_1  |0\rangle +
\frac{1}{2 \pi}({\cal L}_0+{\cal L}_0^\dag) \tilde c_1 |0\rangle +
\frac{1}{2 \pi} ({\cal B}_0+{\cal
B}_0^\dag) \tilde c_0 \tilde c_1 |0\rangle + \frac{\pi}{48} \tilde c_{-1}|0\rangle \nonumber \\
\label{schleve2} &+&\frac{1}{24 \pi}({\cal L}_0+{\cal L}_0^\dag)^2
\tilde c_1|0\rangle +\frac{1}{12 \pi} ({\cal L}_0+{\cal L}_0^\dag)
({\cal B}_0+{\cal B}_0^\dag) \tilde c_0 \tilde c_1 |0\rangle \, .
\end{eqnarray}
This solution was found by solving the string field equation of
motion \cite{Schnabl:2005gv}. In general, a solution to the
equation of motion corresponds to extremal configurations. We are
not guaranteed that the solution will correspond to a minimum,
maximum or saddle point configuration of the tachyon potential.
Certainly, as we have seen in the previous subsection, the
solution lies on the minimum configuration of the effective
tachyon potential. Next we  are going to check whether Schnabl's
solution is a saddle point configuration of the full tachyon
potential\footnote{By the full tachyon potential we mean the
tachyon potential without integrating out the coefficients ($x_i$,
$i \geq 1$).}.

By using Fermat's theorem, the potential extremums of a
multivariable function $f(x_0,\cdots,x_N)$, with partial
derivative $\partial_i f \equiv \frac{\partial }{\partial x_i}
f(x_0,\cdots,x_N)$, $i=0,\cdots,N$ are found by solving an
equation in $\partial_i f=0$. Fermat's theorem gives only a
necessary condition for extreme function values, and some
stationary points are saddle points (not a maximum or minimum). A
test that can be applied at a critical point
$x\equiv(x_0,\cdots,x_N)$ is by using the Hessian matrix, which is
defined as $H_{ij}\equiv\partial_i \partial_j f$. If the Hessian
is positive definite at $x$, then $f$ attains a local minimum at
$x$. If the Hessian is negative definite at $x$, then $f$ attains
a local maximum at $x$. If the Hessian has both positive and
negative eigenvalues then $x$ is a saddle point for $f$.

We claim that Schnabl's analytic solution corresponds to a saddle
point configuration of the full tachyon potential. Evidence
supporting our claim is found by computing the string field
corresponding to the extremal points\footnote{Let us emphasize
that these extremal points are the points corresponding to the
minimum configuration of the effective tachyon potential.}
obtained from extremizing the full tachyon potential, and by
computing the respective eigenvalues of the Hessian matrix. We
have performed this analysis up to level (5,15). The results are
shown in tables \ref{results1} and \ref{results2}. In the first
table \ref{results1} we have compared the first six coefficients
of the analytical $\mathcal{L}_0$ level expansion of the solution
(\ref{schleve2}) with those obtained from extremizing the full
tachyon potential. In the second table \ref{results2}, we show the
respective eigenvalues of the Hessian matrix. It seems that some
eigenvalues of the Hessian matrix does not have pattern of
convergence when the level is increasing. We should attribute the
origin of this divergence to the fact that Schnabl's analytic
solution when expanded in the new bases of $\mathcal{L}_0$
eigenstates results in an asymptotic expansion
\cite{Erler:2009uj}. This issue is in analogy with the problem of
computing the depth of the effective tachyon potential at higher
levels. As we already pointed out, the depth of the effective
potential should diverge since the minimum of the effective
potential should correspond to the value of the string field
action evaluated on Schnabl's solution truncated at some finite
$\mathcal{L}_0$ level.
\begin{table}[ht]
\caption{The six first coefficients of the string field level
expansion corresponding to the saddle point configuration of the
full tachyon potential computed up to level (5,15). The last
column shows the corresponding analytical value for these
coefficients, taken from Schnabl's solution expanded in the
$\mathcal{L}_0$ basis (\ref{schleve2}).} \centering
\begin{tabular}{|c|c|c|c|c|}
\hline
 Level (2,6) & Level (3,9) &  Level (4,12)  &  Level (5,15)  & Coeff. Schnabl's solution   \\
    \hline
0.702361217 & 0.629070893   &  0.632254043 &  0.636801863 & 0.636619772\\
0.165917159 & 0.163408594 & 0.169249348 & 0.160355988 &  0.159154943 \\
0.165917159 & 0.163408594   & 0.169249348 & 0.160355988 & 0.159154943\\
0.036787327 & 0.093717432 & 0.102813347 & 0.064453673 & 0.065449846\\
0.044922378 & 0.005498242  & 0.002281681 & 0.011760238 & 0.013262911\\
0.089844757 & 0.010996484 & 0.004563363 & 0.023520476 &  0.026525823\\
\hline
\end{tabular}
\label{results1}
\end{table}

\begin{table}[ht]
\caption{Eigenvalues of the Hessian matrix corresponding to the
extremal points of the full tachyon potential at different
levels.} \centering
\begin{tabular}{|c|c|}
\hline
 Levels $(m,n)$ & Eigenvalues of the Hessian matrix  \\
    \hline
(2,6) & 327.495421, 223.737479, $-$35.453340, 11.654594   \\
\hline (3,9) & $-$1695.539743, 918.029302, 368.749341,
$-$327.871269\\
& 97.186964, $-$16.995283, 13.192470  \\
\hline  & $-$26154.959292, $-$13583.490319, 5971.278019,
503.332849\\ (4,12)& $-$384.826805, 295.055986, $-$64.304441,
62.708283\\ & 16.044610, $-$14.620229, 6.552447   \\
\hline  & $-$1.445596$\times 10^6$, $\,-$333771.902128,
$\,$226359.468852, $\,$14180.406985 \\ (5,15)& 4335.817536,
$-$2974.472506, 774.905979, $-$680.652950, 5.811130 \\&
445.004921, 255.059728, $-$142.316817, 56.987331\\& $-$16.017131,
$-$11.715162, $-$5.517987, 2.755357\\
\hline
\end{tabular}
\label{results2}
\end{table}

\section{Summary and discussion}
We have given in detail a prescription for computing the tachyon
potential in the sliver frame. As we have seen, calculations are
performed more easily in this frame than in the usual Virasoro
basis of $L_0$ eigenstates. For instance, in the old basis the
evaluation of the cubic interaction term using CFT methods
requires cumbersome computations of finite conformal
transformations for non-primary fields. The simplicity of the
definition of the $*$-product in the new basis allows us to
overcome these difficulties.

Since one aim of this paper was to answer the question whether
Schnabl's analytic solution corresponds to a saddle point
configuration of the full tachyon potential, we have focused our
attention to a string field satisfying the Schnabl gauge. Since
the computation of the tachyon potential does not require us to
choose a specific gauge condition, we can use another gauge for
the string field. It would be interesting to find connections
between those family of solutions computed in different gauges
which gives the right value for the vacuum energy. For instance,
in recent work \cite{Erler:2009uj} a new simple solution for the
tachyon condensation was analyzed and an explicit gauge
transformation which connects the new solution to the original
Schnabl's solution was constructed \cite{Schnabl:2005gv}.

We have provided a strong evidence that Schnabl's analytic
solution corresponds to a saddle point configuration of the full
tachyon potential, and furthermore we have shown that the solution
lies on the minimum of the effective tachyon potential.
Nevertheless, there remain two important issues regarding the
vacuum solution. The first is related to the computation of the
analytic solution in the Siegel gauge. The second is to construct
an explicit gauge transformation which connects Schnabl's solution
to the one found using the Siegel gauge.

An issue that could be addressed using the methods outlined in
this work would be the computation of the effective tachyon
potential in the sliver frame for the case of the cubic
superstring field theory. The profile of the effective potential
in this theory is very puzzling since the tachyon has vanishing
expectation value at the local minimum of the effective potential,
so the tachyon vacuum sits directly below the perturbative vacuum
\cite{Erler:2007xt}.

Finally, while we have not developed the details here, our methods
should be applicable to computations in Berkovits's superstring
field theory. The relevant string field theory is non-polynomial
\cite{Berkovits:1995ab}, but since the theory is based on Witten's
associative star product, the methods discussed in this paper
would apply with minor modifications. It would certainly be
desirable to test the brane-antibrane annihilation conjecture
analytically \cite{aldo3}.

\section*{Acknowledgements}
I would like to thank Nathan Berkovits, Ted Erler, Michael Kroyter
and Martin Schnabl for useful discussions. I also wish to thank
Diany Ruby, who proofread the manuscript. This work is supported
by CNPq grant 150051/2009-3.





\end{document}